# Structured Image-based Coding for Efficient Gaussian Splatting Compression

Pedro Martin, António Rodrigues, João Ascenso, and Maria Paula Queluz

Instituto de Telecomunicações, Instituto Superior Técnico, University of Lisbon, 1049-001 Lisbon, Portugal
{pedro.martin, antonio.rodrigues, joao.ascenso, paula.queluz}@lx.it.pt

***Abstract*—Gaussian Splatting (GS) has recently emerged as a state-of-the-art representation for radiance fields, combining real-time rendering with high visual fidelity. However, GS models require storing millions of parameters, leading to large file sizes that impair their use in practical multimedia systems. To address this limitation, this paper introduces GS Image-based Compression (GSICO), a novel GS codec that efficiently compresses pre-trained GS models while preserving perceptual fidelity. The core contribution lies in a mapping procedure that arranges GS parameters into structured images, guided by a novel algorithm that enhances spatial coherence. These GS parameter images are then encoded using a conventional image codec. Experimental evaluations on Tanks and Temples, Deep Blending, and Mip-NeRF360 datasets show that GSICO achieves average compression factors of 20.2× with minimal loss in visual quality, as measured by PSNR, SSIM, and LPIPS. Compared with state-of-the-art GS compression methods, the proposed codec consistently yields superior rate-distortion (RD) trade-offs.**

## I. INTRODUCTION

The rise of 3D visual representation has revolutionized multimedia consumption, enabling immersive and interactive experiences. From virtual reality (VR) gaming and virtual museum tours to augmented reality (AR) based surgical procedures for medical training, 3D content has expanded the boundaries of multimedia applications. In recent years, neural radiance fields (NeRFs) [1] have established themselves as a breakthrough paradigm for 3D scene representation and novel view synthesis, offering unprecedented levels of visual realism unattainable with traditional image-based approaches. By modeling the volumetric scene as a radiance function using neural networks (NNs) trained to map spatial coordinates and viewing directions to color and opacity values, NeRF achieves compelling results across synthetic and real-world datasets. However, the computational cost associated with NeRF training and rendering pose critical challenges for practical deployment. To address these challenges, several extensions and alternative representations have been introduced, with particular emphasis on improving training and rendering efficiency while preserving high-quality visual synthesis. Gaussian Splatting (GS) [2] has recently emerged as one of the most promising alternatives to NeRF. Instead of encoding a scene into a neural volumetric function, GS represents 3D content using a set of anisotropic 3D Gaussian primitives, parameterized by spatial position, scale, rotation, opacity, and spherical harmonic (SH) coefficients for view-dependent color representation. Rendering is performed via a tile-based rasterization process, avoiding the expensive ray-marching approach used in NeRF. This design allows GS to achieve real-time frame rates, while reducing training times and avoiding unnecessary computations in empty 3D space. The seminal 3D Gaussian Splatting (3DGS) method [2], has become a new baseline in radiance field research, inspiring a growing number of variants. GS representations have already been applied to several domains, including immersive VR/AR applications, telepresence, cultural heritage preservation, and large-scale scene reconstruction [3].

Although GS methods offer significant advantages, they demand large amounts of data to represent a scene. A single GS model size may achieve hundreds of megabytes, limiting its use in bandwidth-constrained or storage-limited settings. Existing GS compression methods have primarily combined pruning, quantization, and entropy modeling within joint training-compression pipelines. However, such approaches often require retraining or fine-tuning the model, which may not be feasible in many real-world scenarios where the original images or videos used to train the model are no longer available. In fact, in many practical settings, the creator of the 3D content is not the party responsible for preparing and compressing it for distribution. This paper addresses such cases, where compression is applied after the 3D model has already been generated (i.e., post-training), by proposing a solution that is broadly applicable and agnostic to the underlying GS model.

In this context, the main objective of this work is to provide an efficient and versatile solution for compressing GS models without requiring any post-training optimization or fine-tuning of the model. To validate the approach, two widely-used GS baselines are considered, 3DGS [2] and Scaffold-GS [4], demonstrating wide applicability. The proposed compression solution, referred to as GS Image-based Compression (GSICO), converts GS parameters into structured image representations, which are then compressed by a conventional image codec. To further improve compressibility, the design incorporates a dedicated GS mapping algorithm, using a novel Nearest-Neighbor-based Sorting (NNS) that increases spatial coherence within the parameter images, prior to encoding. This image-based paradigm not only leverages decades of progress in image compression, but also embodies the novelty of treating GS compression through a codec-like perspective; this contrasts with existing GS compression methods that rely heavily on retraining or parameter fine-tuning. The main contributions of this paper can be summarized as follows:

- **Novel GS Codec:** The proposed GSICO codec introduces a novel training-free and model-agnostic compression framework capable of handling quite different GS representations, such as 3DGS and Scaffold-GS. Operating strictly post-training, it remains independent of any model optimization stage, enabling seamless integration into a wide range of GS pipelines. Extensive evaluation on widely used radiance field datasets and comparison with state-of-the-art GS compression benchmarks demonstrates its effectiveness. GSICO achieves average compression factors of 20.2× for 3DGS inputs and 8.5× for Scaffold-GS inputs, with negligible degradation in rendering quality.
- **2D GS Mapping Algorithm:** A second key contribution is the introduction of a novel parameter mapping algorithm, that clusters and spatially arrange GS parameters in the 2D image domain (using a novel NNS algorithm), to improve local coherence. This strategy substantially reduces entropy, thereby improving compression efficiency when paired with a standard image codec. Furthermore, NNS enforces consistent Gaussian-to-pixel alignment across all parameter maps, ensuring coherent and fast reconstruction during decoding.

The link for the GSICO implementation will be available after acceptance.

## II. Related Work

Research on radiance field representations and their compression has evolved rapidly in recent years. This section reviews the most relevant prior work, focusing on three main directions: *i)* compression of NeRF-based representations, *ii)* compression of GS-based representations, and *iii)* current challenges in GS compression.

### *A. NeRF Representation and Compression*

NeRF [1] established a new paradigm for novel view synthesis by learning implicit volumetric scene representations using multilayer perceptrons (MLPs). While NeRF methods achieve highly realistic view synthesis, their dependence on dense ray sampling and large NNs makes both training and rendering computationally demanding. To attenuate these costs, several explicit extensions have been proposed, including hash-based encodings, voxel grids, and tensor factorization [5]-[7]. These designs significantly accelerate rendering but often lead to large model sizes. Consequently, compression of NeRF representations has become an active research direction. Early works employed parameter pruning, vector quantization, and entropy coding to reduce the NeRF model size [8], [9], while more recent approaches explored transforms such as Fourier and wavelet bases [10], [11].

### *B. GS Representation and Compression*

Another step forward in radiance field representations was the introduction of the GS representation. The seminal 3DGS method [2], explicitly models a scene using millions of 3D Gaussians, each parametrized by its position, $\mathbf{x}_\mathrm{G} = (x_G, y_G, z_G)$, scale, $\mathbf{s} = (s_x, s_y, s_z)$, rotation (defined by the quaternion representation), $\mathbf{r} = (r_u, r_x, r_y, r_z)$, opacity, $\sigma$, and SH (for view-dependent color), $\mathbf{k} = (k_1, \dots, k_{48})$. Rendering is performed via differentiable rasterization, enabling real-time visualization and substantially faster training than NeRF, while preserving high visual quality. However, the cost of storing millions of Gaussians, each described by several parameters (many of them with more than one element), results in a large amount of data posing practical challenges for storage and transmission. Scaffold-GS [4] mitigates this limitation by introducing a novel GS representation built on voxel anchors. Instead of storing each Gaussian explicitly, the scene is represented using a sparse set of voxels distributed in 3D space. Each voxel has a learnable anchor whose position, $\mathbf{x}_\mathrm{A} = (x_A, y_A, z_A)$, defines its spatial location, and a scale factor, $S_f$, determines the voxel size, allowing the representation to adapt to local geometric complexity. Multiple neural Gaussians are associated to each anchor, to model fine-grained geometry and appearance within that anchor voxel. These Gaussians are called neural because their attributes are not stored explicitly but predicted through small MLPs. The spatial position of each neural Gaussian is computed from a vector of learned offset features, $\mathbf{O} = (O_1, \dots, O_{30})$, that encodes displacements relative to the anchor position. Each voxel also includes a vector of anchor features, $\mathbf{A} = (A_1, \dots, A_{32})$, that encodes the geometry and appearance of its associated neural Gaussians. Small MLPs take as input the offset and anchor features to predict each neural Gaussian scale, rotation, opacity, and color. This two-level organization forms a hierarchical structure: anchors represent the coarse information of the scene, while the neural Gaussians represent fine scene details. As a result, Scaffold-GS achieves a more compact and efficient scene representation than 3DGS. Fig. 1 illustrates the contrast between 3DGS, which uses individual and independent Gaussians, and Scaffold-GS, which groups them around latent anchors. Together, 3DGS and Scaffold-GS represent the two dominant architectures in GS.

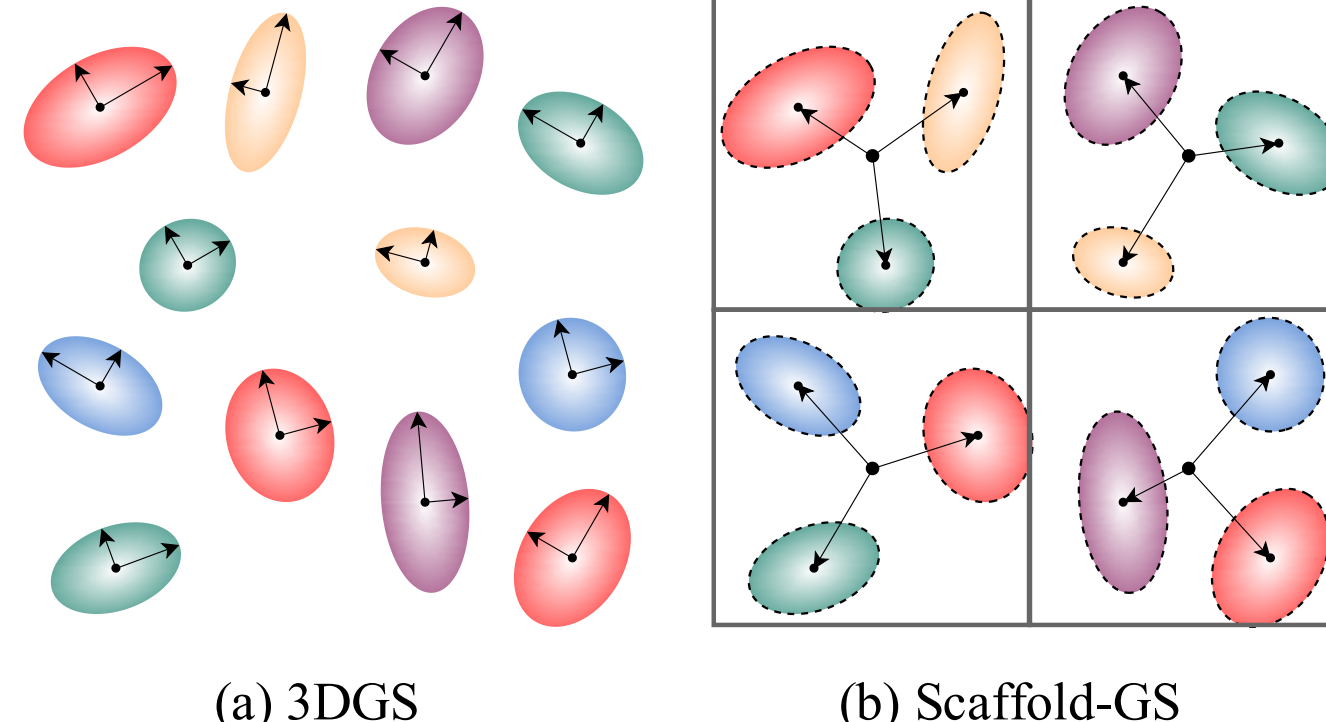

**Fig. 1:** Illustration of the model representation of the 3DGS and Scaffold-GS methods.

Research on GS compression can be broadly categorized into three strategies: *i)* pruning-based techniques that reduce model size by discarding redundant Gaussians or SH coefficients [12]-[15]. Pruning is typically guided by heuristics or learned metrics that evaluate each Gaussian perceptual relevance. While pruning can achieve substantial reductions, aggressive removal may considerably degrade the rendering quality; *ii)* quantization and codebook methods, that represent GS parameters with a limited set of representative symbols (codewords) through some mapping [12]-[14], [16]-[18]. This significantly reduces the model size and can achieve high compression ratios with limited perceptual loss, but requires careful codebook design to avoid visual artifacts; and *iii)* entropy modeling, that exploits statistical redundancies in Gaussian parameters [12]-[14], [19]. By learning probability distributions across GS parameters, entropy-based methods can produce compact bitstream sizes.

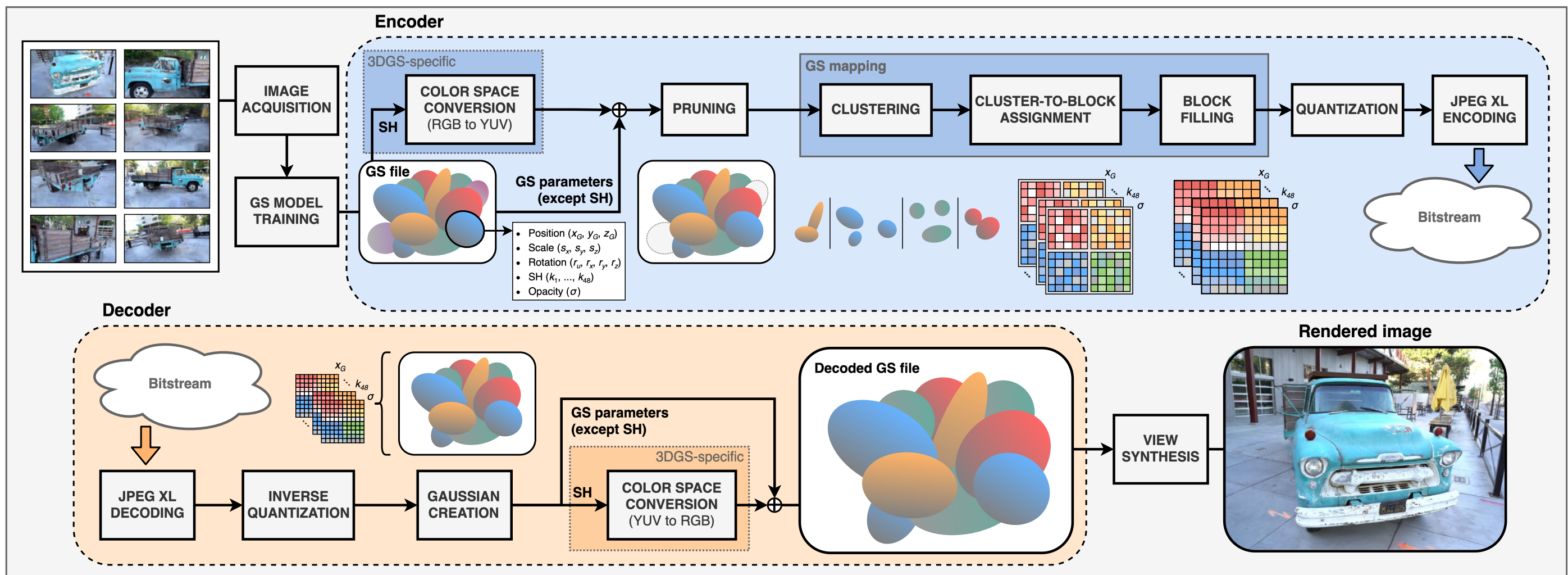

**Fig. 2:** GSICO encoder (blue) and decoder (orange) framework pipeline.

### C. Limitations of Existing GS Compression Methods

Despite their effectiveness, most existing GS compression methods exhibit a major limitation; they are training-dependent, requiring either full retraining or fine-tuning after compression. This severely constrains their applicability, since it is necessary to have the original images used for training, and significant computational resources to compress the model. In many areas (e.g., point cloud compression), the model is traditionally obtained first and compressed as it is, maintaining as much fidelity to the original model as possible. In addition, many compression methods are tailored to a specific GS representation (e.g., 3DGS or Scaffold-GS), resulting in limited use across different baseline models.

A promising yet underexplored direction is to leverage advanced image coding technologies by mapping Gaussian parameters into 2D layouts compatible with standard image/video codecs [15], [16]. By exploiting spatial correlations through established image/video codecs, such strategy can achieve substantial reductions while avoiding the need for retraining or fine-tuning. Actually, this was explored for other 3D representations such as point clouds (e.g., within the MPEG V-PCC standard [20]). Image/video codecs bring additional advantages: they benefit from decades of optimization, have standardized decoders, and are hardware-accelerated on most platforms. Further exploration of this direction is crucial to bridge the gap between GS compression research and broader multimedia coding systems. The proposed GSICO codec directly addresses this need by introducing a compression strategy compatible with both 3DGS and Scaffold-GS, the two currently most representative GS model baselines.

## III. GSICO Framework and Modules

This section presents a comprehensive overview of the GSICO framework, detailing each module of the encoder and decoder pipelines. Fig. 2 illustrates the overall processing flow, whose architecture supports both 3DGS-based and Scaffold-GS-based representations (except for the color-space conversion module, which is specific to 3DGS). The color figures shown beneath each module correspond to the 3DGS representation and are included to offer an intuitive visual interpretation of how GSICO operates on explicit Gaussian primitives.

### A. Walkthrough

The GSICO encoder accepts models produced by both 3DGS-based and Scaffold-GS-based methods, offering flexibility and broad applicability. Table I summarizes the parameters associated with each model. Although Scaffold-GS introduces additional model specific parameters, it generally requires far fewer voxels than the number of Gaussians used in 3DGS, leading to a more compact and efficient representation.

3DGS-based models store, for each Gaussian, its position, rotation, scale, opacity, and SH coefficients. The latter spans four degrees (0 to 3), resulting in 16 coefficients per RGB color channel: one DC coefficient (for degree-0), and three, five, and seven AC coefficients, for degrees 1, 2, and 3, respectively. SH are a set of base functions defined on the sphere surface and, in the context of GS, are used to efficiently model view-dependent color variations for each Gaussian. Their perceptual contribution decreases with increasing degree, as higher-order coefficients correspond to higher frequency variations, analogous to the role of high-frequency components in DCT-based transforms.

TABLE I
GS MODEL REPRESENTATION IN THE GS FILE DATA STRUCTURE

(a) 3DGS

| GS parameter | #param per Gaussian |
|---|---|
| Position, $(x, y, z)$ | 3 |
| SH DC, $(k_1, k_2, k_3)$ | 3 |
| SH AC, $(k_4, \ldots, k_{48})$ | 45 |
| Scale, $(s_x, s_y, s_z)$ | 3 |
| Rotation, $(r_u, r_x, r_y, r_z)$ | 4 |
| Opacity, $(\sigma)$ | 1 |

(b) Scaffold-GS

| GS parameter | #param per voxel |
|---|---|
| Position (anchor), $(x, y, z)$ | 3 |
| Offset features, $(O_1, \ldots, O_{30})$ | 30 |
| Anchor features, $(A_1, \ldots, A_{32})$ | 32 |
| Scale factor, $(S_f)$ | 1 |

Scaffold-GS-based models store, for each voxel, anchor positions, scale factor, offset and anchor features, and three small MLPs that predict Gaussian attributes. Although MLP parameters must also be transmitted, their size is relatively small (7086 weights and biases, in total), when compared to the remaining model parameters. For this reason, GSICO does not compressed them; instead, they are transmitted directly in the bitstream (though a standard like MPEG NNR [21] could be used instead).

The key idea behind GSICO is to convert the rich 3D representation used in GS into a series of 2D parameter maps (or a structured 3D volume), where each map (or slice) stores one parameter of the GS model. Under this approach, the number of maps equals the number of parameters, $P$ (with $P = 59$ for 3DGS and $P = 66$ for Scaffold-GS), and the total number of pixels of each map matches the number of Gaussians or voxels, in the underlying 3D representation. As shown in Fig. 3, in red color, a given pixel location, $(x, y)$, is consistently associated with the same Gaussian (or voxel) across all parameter maps. In this sense, $(x, y)$ acts as a lookup index that retrieves the full set of GS parameters stored across the maps.

Simply arranging GS parameters into 2D maps is not sufficient to achieve efficient compression. Each map stores parameters from different Gaussians (or voxels) with limited correlation between them. To enable efficient compression with a conventional image/video codec, spatial redundancy within each map must be exploited. For this purpose, a novel strategy is introduced to generate the structured 3D volume: Gaussians (or voxels) are first grouped into fixed-size clusters based on similarity and then assigned to 3D blocks; each block is filled with the Gaussians (or voxels) parameters of one cluster, as illustrated with orange color in Fig. 3. A dedicated similarity criterion, specifically designed for this task, guides this process. This criterion relies on SH coefficients (for 3DGS-based inputs), as these dominate the data volume, or on anchor positions and offset features (for Scaffold-GS-based inputs), as this combination empirically showed to correlate well with compression performance. This approach enhances local similarity within each parameter map, enabling significantly more efficient compression than naïve approaches such as random ordering. The GSICO encoder modules are described next:

- **Color space conversion:** The SH coefficients represent the view-dependent color of each Gaussian and are originally defined in the RGB color space. To reduce redundancy across color components, the linear RGB-to-YUV transform specified in ITU-R BT.601 [22] (used without any clipping) is directly applied to the SH coefficients. This concentrates most of the signal energy in the luminance component, enabling the pruning of chrominance AC SH coefficients with negligible perceptual impact. As both the RGB-to-YUV transform and the subsequent GS mapping of SH coefficients are linear operations, applying the color transform before (in the pixel domain) or after (in the SH coefficients domain) the GS mapping is mathematically equivalent. However, since GS mapping uses the SH coefficients within its similarity criterion, applying the color transform beforehand ensures that it operates on already decorrelated color components, improving its effectiveness. This color space transform operation is applied only to 3DGS-based inputs, since Scaffold-GS does not employ SH-based color parametrization.
- **Pruning:** This step ensures that Gaussian (or voxel) parameters can be organized into rectangular-shaped 2D maps, while slightly reducing the data volume and preserving rendering quality. Since the resulting maps are composed of 16×16 blocks (the block size used in the next steps), both width and height must be multiples of 16 pixels. The chosen block size aligns with those commonly employed in standard image codecs, facilitating better compatibility with the subsequent image coding step. Other common block sizes (such as 4×4, 8×8, 32×32, 64×64, and 128×128 pixels) were evaluated but did not provide a better trade-off between computational efficiency and compression performance. In particular, larger block sizes yield only marginal compression gains while significantly increasing runtime. The width and height map sizes, $W_{map}$ and $H_{map}$, respectively, are determined based on the total number of blocks that can be formed by all Gaussians (or voxels). Thus, if the total number of Gaussians (or voxels) is not divisible by 16, the excess elements are pruned to satisfy this requirement. Furthermore, if the resulting number of blocks that can be formed from all Gaussians (or voxels) is a prime number, 256 Gaussians (equivalent to one 16×16 block) are pruned to make the block count factorable. The final width and height are then chosen as the two closest integer factors (both multiples of 16) whose product matches the remaining number of elements. Whenever pruning is required, opacity is used as the selection criterion, as it has proven to be a reliable indicator of the visual importance of each Gaussian [15]. Gaussians with lower opacities (or voxels with lower average Gaussian opacities) contribute less to the rendered quality, making them suitable candidates for pruning.

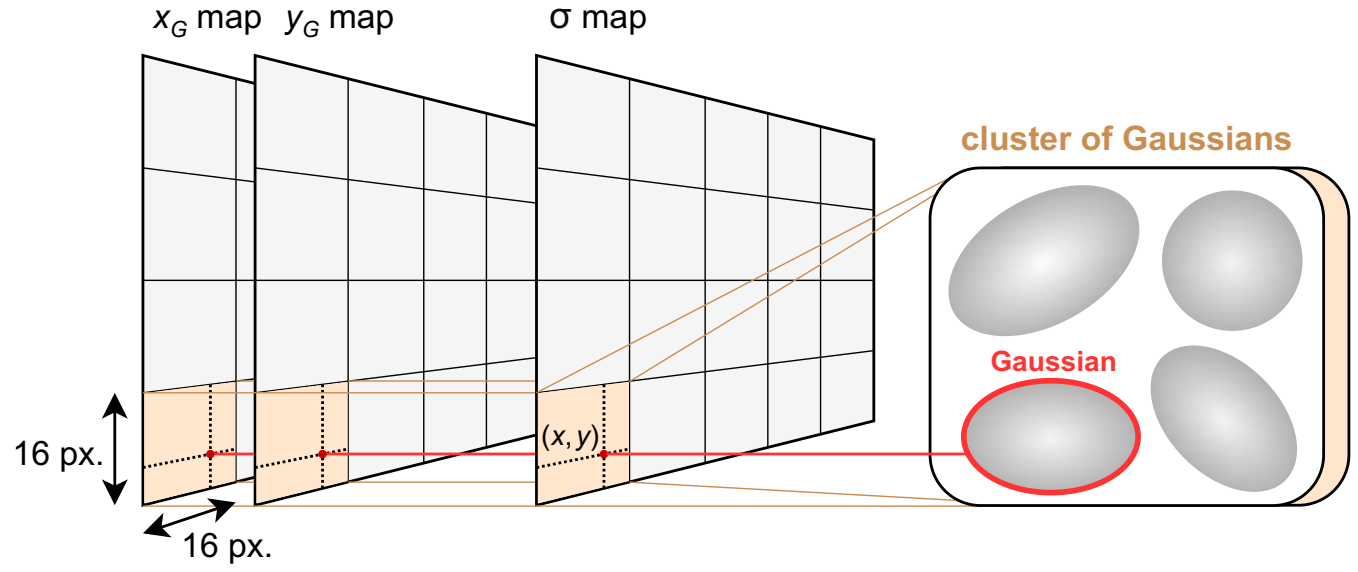

(a) 3DGS-based input files

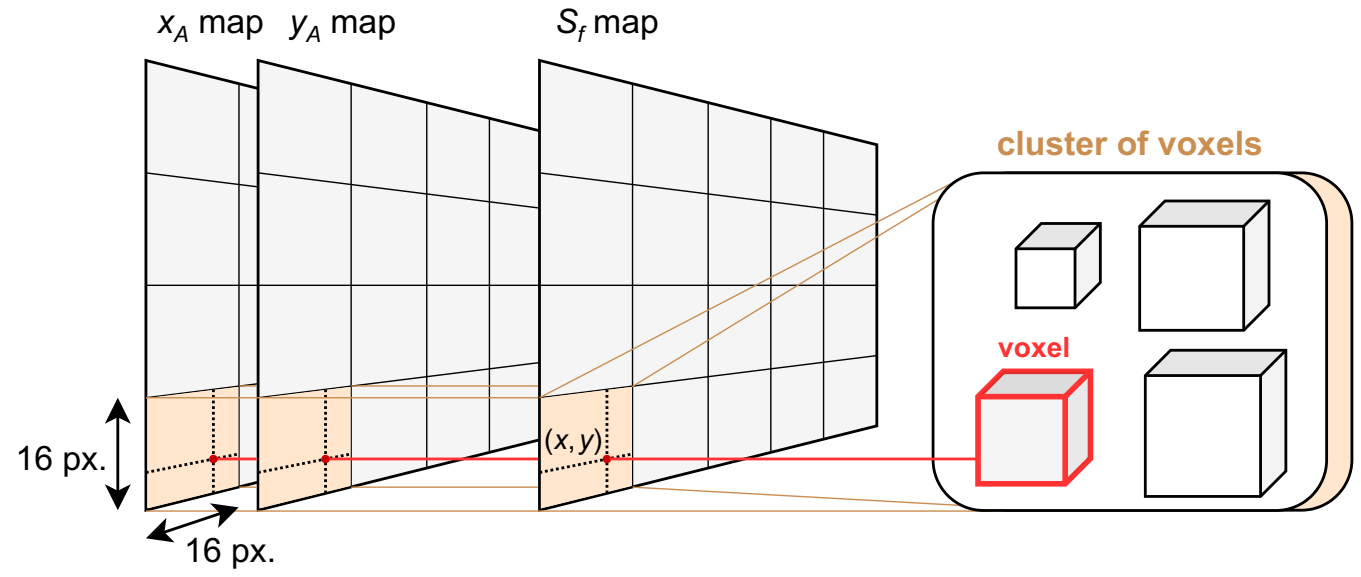

(b) Scaffold-GS-based input files

**Fig. 3:** Illustration of the GS parameter mapping for the 3DGS-based and Scaffold-GS-based input files.

- **Clustering:** In this step, Gaussians (or voxels) are grouped into fixed-size clusters to enable the block-based organization of their parameters in the 2D maps. Each cluster contains 256 elements, allowing their parameters to be later arranged as a 16×16 block on the corresponding map (cf. Fig. 3). Clustering is performed in a feature space (using a similarity criterion) defined by GS parameters that were found most relevant to compression efficiency. The specific feature space depends on the baseline model: for 3DGS-based inputs, clustering relies on the luminance SH AC coefficients, while for Scaffold-GS-based inputs it consists of anchor positions concatenated with offset features. This step ensures that Gaussians (or voxels) with similar parameters are placed together in the same block, increasing spatial coherence within each parameter map. Additional implementation details are provided in Section III.B.
- **Cluster-to-block assignment:** This step receives as input the set of clusters computed in the previous step and assigns each one to a 3D block on the set of 2D maps (or 3D volume). The goal is to place clusters with similar characteristics near one another, improving global spatial coherence; the actual placement of parameter values within each block is performed in the next step. A novel NNS algorithm defines the association between clusters and blocks, using the same similarity criterion employed during clustering. For this purpose, each cluster is represented by a feature vector obtained by averaging the relevant parameters (namely, the luminance SH AC coefficients for 3DGS-based inputs or the anchor positions concatenated with offset features for Scaffold-GS inputs) across all elements in that cluster. Additional implementation details are provided in Section III.C.
- **Block filling:** This step takes as input the clusters and their assigned block locations, and fills each 3D block with the corresponding GS parameter values. Note that each slice of the 3D block corresponds to one parameter. This filling is done block-by-block in snake scan order, starting with the top-left block and proceeding horizontally. The same NNS algorithm used for cluster-to-block assignment is now applied for block filling to determine the placement of each Gaussian (or voxel) within the 3D block, using the same similarity criterion of the clustering step. This enforces that neighboring positions on a given 2D map correspond to elements with similar parameter values, thereby enhancing local spatial coherence. Further details are provided in Section III.C.
- **Quantization:** This step receives as input the set of 2D maps produced previously, which are in floating-point, and applies uniform mid-tread quantization. The goal is to reduce parameter precision, thereby lowering the bitrate after image compression while preserving the quality of the reconstructed scene. Furthermore, for 3DGS models, the SH AC chrominance components are discarded to further reduce the bitrate, with negligible perceptual impact. For quantization, each $i$-th parameter (and thus each 2D map) is assigned a fixed bit depth, $b_i$, determined empirically based on extensive experiments. Since the value ranges differ across parameters, the quantization step size is computed independently for each map based on the minimum, $x_{\min}$, and maximum, $x_{\max}$, observed values. The quantization step size is then given by:

$$Q_{\text{step}}^{i} = \frac{x_{\max}^{i} - x_{\min}^{i}}{2^{b_i}} \quad (1)$$

Both the minimum and maximum values of each parameter are included in the bitstream, enabling the decoder to compute the step size, $Q_{\text{step}}^{i}$. The bit depth assigned to each parameter is assumed to be known to both the encoder and decoder. Finally, each 2D map is independently quantized according to:

$$Q_{\text{index}}^{i} = \text{round}\left(\frac{x_i - x_{\min}^{i}}{Q_{\text{step}}^{i}}\right) \quad (2)$$

- **JPEG XL encoding:** This step encodes the set of quantized 2D maps from the previous step using the JPEG XL image codec [23]. JPEG XL was selected for its high compression efficiency, especially for structured data such as the GS parameters maps. JPEG XL provides a quality parameter that regulates the balance between compression efficiency and fidelity, ranging from negative infinity to 100, where 100 denotes lossless coding. For 3DGS-based models, the SH parameter maps are encoded in lossy mode, while all other parameter maps are encoded in lossless mode. This leverages the lower sensitivity of SH coefficients to small distortions, while preserving geometric parameters. In its lossy configuration, JPEG XL employs 8×8 DCT-based predictive coding. As discussed earlier, this characteristic influenced the selection of the block size used for GS parameter maps creation. For Scaffold-GS-based models, all maps are encoded using JPEG XL in lossless mode. This choice reflects the voxel-based nature of Scaffold-GS, where even small losses introduced by image compression can propagate and result in noticeable rendering artifacts. Since each parameter corresponds to a dedicated 2D map, encoding is performed on a map-by-map basis, with each map treated as a single-channel luminance image. This step constitutes the final stage of the GSICO encoder. The resulting bitstream includes the JPEG XL-encoded 2D maps along with the associated quantization metadata (minimum and maximum parameter values), and, in Scaffold-GS-based models, also includes the MLP parameters.

The GSICO decoder performs the inverse operations of the encoder to reconstruct a GS file from the codec bitstream. It first decodes the compressed GS parameter maps, after which inverse quantization is applied to get their floating-point values. Specifically, each integer index is mapped back to its reconstruction level using $x'_i = Q^i_{\text{index}} Q^i_{\text{step}} + x^i_{\min}$ , where the step size is derived from the minimum and maximum parameter values transmitted in the bitstream. Finally, the Gaussians (or voxels) are reconstructed by reading the same pixel position across all reconstructed 2D maps, since each position across the parameter images corresponds to a single Gaussian (or voxel). This synchronized organization enables fast decoding. For Scaffold-GS-based models, this is the last process of GSICO, enabling the restored file to be used directly for rendering novel views. For 3DGS-based models, the inverse color space conversion (from YUV-to-RGB) is further applied to the SH coefficients. At the end of the codec pipeline, the decoded GS file can be used for view synthesis from camera positions requested by a user.

The GSICO operating points were selected empirically by assessing how quantization and JPEG XL coding impact the rendering quality, leading to the configuration presented in Table II. For 3DGS-based inputs, different quantization settings are applied to the SH coefficients; therefore, Table II distinguishes between luminance (Y) and chrominance (UV) coefficients, as well as between first, second, and third-degree SH coefficients (1º, 2º, and 3º). Five operating points were defined for each representation, namely 3DGS-based and Scaffold-GS-based, ensuring that the resulting RD curves reflect a meaningful trade-off between rate and quality. For 3DGS, the operating points are most effectively tuned by varying the JPEG XL quality level in lossy mode, as this provides the dominant RD behavior. In contrast, for Scaffold-GS (where the voxel structure is more sensitive to compression), JPEG XL is used exclusively in lossless mode, and the operating points are instead determined by varying the quantization bit depth configuration.

While the overview above outlines the full GSICO pipeline, its core contribution lies in the GS mapping encoder, which organizes GS parameters into a set of 2D maps in a compression-efficient manner. Section III.B and Section III.C detail the key modules involved in this process.

### *B. Clustering*

A random mapping of Gaussians (or voxels) parameters onto 2D maps (or 3D volume) is inefficient, since it leads to very limited small spatial correlation and therefore poor coding efficiency. On the other hand, exhaustively testing all possible combinations of Gaussians (or voxels) parameters would be computationally expensive and impractical due to the typically large number of elements involved (in the order of millions). Clustering provides an efficient intermediate step by first grouping similar elements, that can later be organized into 3D blocks at much lower computational cost.

The key idea is to perform fixed-size clustering of Gaussians (or voxels), grouping them according to the similarity of their parameter values. A fixed cluster size allows the resulting data to be arranged into 2D maps using a regular grid of 16×16-pixel blocks, which aligns well with the design characteristics of standard image/video codecs.

Given its optimality in minimizing intra-cluster variance, the K-Means algorithm [24] was adopted as the basis of the clustering process. Since K-Means does not inherently support fixed-size clusters, a K-Means-based fixed-size clustering algorithm was developed. This algorithm receives as input a set of $N$ Gaussians (or voxels) with the respective parameter values, $\mathbf{G} = \{\mathbf{g}_1, \ldots, \mathbf{g}_N\}$ and outputs a set of $L$ fixed-size clusters of 256 elements (to enable the creation of 16×16-pixel blocks), $\mathbf{C} = \{\mathbf{C}_1, \ldots, \mathbf{C}_L\}$, with $L = N/256$. The pruning step already ensured that the total number of Gaussians (or voxels) to be clustered are multiple of 256. The clustering algorithm consists of five steps:

1. **Initialization:** Initialize $\mathbf{C}$ to an empty set;
2. **Clustering:** Apply K-Means to $\mathbf{G}$, with K-Means++ initialization, to obtain $K = \text{card}(\mathbf{G})/256$ intermediate clusters (initially, $K = L$), $\mathbf{C_I} = \{\mathbf{C}^\mathbf{I}_1, \ldots, \mathbf{C}^\mathbf{I}_K\}$. In K-Means, similarity is measured using the Euclidean distance:

$$d\big(\mathbf{f}(\mathbf{g}_i), \mathbf{c}_j\big) = \left\|\mathbf{f}(\mathbf{g}_i) - \mathbf{c}_j\right\|_2, \tag{3}$$

where $\mathbf{f}(\mathbf{g}_i)$ denotes the feature vector of the $i$-th element from $\mathbf{G}$ and $\mathbf{c}_j$ is the centroid of the $j$-th cluster. For

TABLE II
GSICO OPERATING POINTS DEFINED BY THE QUANTIZATION BIT DEPTH AND JPEG XL QUALITY LEVEL

(a) 3DGS-based input files

| | RD point | GS parameters | | | | | | | |
|---|---|---|---|---|---|---|---|---|---|
| | | **P1** | **P2** | **P3** | **P4** | **P5** | **P6** | **P7** | **P8** |
| **Quantization** (bit depth) | **All** | 14 | 8 | 8 | 8 | 6 | 5 | 0 | 6 |
| **JPEG XL** (quality level) | **1** | 100 | 100 | 100 | 96 | 96 | 96 | 96 | 100 |
| | **2** | 100 | 100 | 100 | 90 | 90 | 90 | 90 | 100 |
| | **3** | 100 | 100 | 100 | 70 | 70 | 70 | 70 | 100 |
| | **4** | 100 | 100 | 100 | 20 | 20 | 20 | 20 | 100 |
| | **5** | 100 | 100 | 100 | 0 | 0 | 0 | 0 | 100 |

(P1 - position, P2 - scale, P3 - rotation, P4 - SH DC, P5 - SH AC Y 1º, P6 - SH AC Y 2º/3º, P7 - SH AC UV, P8 – opacity)

(b) Scaffold-GS-based input files

| | RD point | GS parameters | | | |
|---|---|---|---|---|---|
| | | **P1** | **P2** | **P3** | **P4** |
| **Quantization** (bit depth) | **1** | 16 | 8 | 8 | 8 |
| | **2** | 16 | 8 | 8 | 6 |
| | **3** | 16 | 8 | 6 | 6 |
| | **4** | 16 | 8 | 6 | 4 |
| | **5** | 16 | 8 | 4 | 4 |
| **JPEG XL** (quality level) | **All** | 100 | 100 | 100 | 100 |

(P1 - position, P2 - scale factor, P3 - offset features, P4 - anchor features)

3DGS-based inputs, $\mathbf{f}(\cdot)$ consists of the luminance SH AC coefficients; for Scaffold-GS-based inputs, it is formed by the concatenation of anchor positions and offset features. Each centroid $\mathbf{c}_j$ is computed as the element-wise average of the feature vectors in the respective intermediate cluster, $\mathbf{f}_{\text{avg}}(\mathbf{C}_j^{\mathbf{I}})$.

3. **Extract fixed-size clusters:** Each cluster in $\mathbf{C}_{\mathbf{I}}$ with more than 256 elements is randomly partitioned into disjoint sub-clusters of size 256, with each resulting sub-cluster added to $\mathbf{C}$. Other splitting strategies were evaluated without leading to significant differences in performance, as K-Means already groups elements that are similar between them.
4. **Update the remaining set:** Remove from $\mathbf{G}$ the elements assigned to clusters of $\mathbf{C}$ in the previous step. Any leftover elements (resulting from the clusters $\mathbf{C}_j^{\mathbf{I}}$ whose cardinal is not a multiple of 256) remain in $\mathbf{G}$.
5. **Stopping criteria**: If $\mathbf{G} \neq \emptyset$, repeat from the second step using K-Means on the remaining elements of $\mathbf{G}$.

*C. Cluster-to-Block Assignment and Filling*

The fixed-size clusters previously obtained are assigned to blocks within a set of 2D maps (or 3D volume). However, a spatial organization is required (both across blocks and within each block) to ensure that similar clusters and their elements are placed close to each other. This spatial coherence is essential to improve coding performance. To this end, a novel Nearest-Neighbor-based Sorting (NNS) algorithm was developed to spatially organize Gaussians (or voxels), or their clusters, within the set of 2D maps. NNS ensures that elements with similar characteristics are placed at neighboring spatial locations, thereby inducing strong local correlations. This spatial organization leads to smooth variations across the parameter maps, which are well aligned with the prediction and transform mechanisms exploited by standard image codecs. As a consequence, the resulting 2D maps exhibit reduced entropy and can be compressed more efficiently. In the cluster-to-bock assignment procedure, the NNS algorithm is used to determine the placement of clusters within the set of blocks; in the block filling procedure, it defines the placement of the cluster Gaussians (or voxels) inside the block.

The NNS algorithm takes as input a set of $S$ elements, $\mathbf{E} = \{\mathbf{e}_1, \dots, \mathbf{e}_S\}$, and two integers, $W$ and $H$, such that $S = W \times H$; for cluster-to-block assignment, $\mathbf{e}_i$ represents the centroid of a cluster obtained in the clustering step, $W = W_{map}/16$, and $H = H_{map}/16$, where $W_{map}$ and $H_{map}$ are, respectively, the width and height of the 2D maps (computed during the pruning step described in Section III.A); thus, in this case, $S$ corresponds to the total number of clusters (i.e., 16×16 blocks to be formed); for block filling, $\mathbf{e}_i$ represents a Gaussian (or voxel), $W = 16$ and $H = 16$. The output is a 2D matrix, $\mathbf{M}$, with size $W \times H$, that contains in each position the index of an element of $\mathbf{E}$. Accordingly, the NNS algorithm can be defined as $\mathbf{M} = \text{NNS}(\mathbf{E}, W, H)$, being composed of four steps:

1. **Initialization and first assignment:** Compute the element-wise median feature vector of the entire set $\mathbf{E}$, denoted as $\mathbf{f}_{\text{med}}(\mathbf{E})$; the feature vector of each element, $\mathbf{f}(\cdot)$, is the same considered for the clustering step. Select the element $\mathbf{e}_k$ from $\mathbf{E}$ whose feature vector $\mathbf{f}(\mathbf{e}_k)$ is closest to $\mathbf{f}_{\text{med}}(\mathbf{E})$, in terms of Euclidean distance:
$$\mathbf{e}_k = \arg\min_{\mathbf{e}_l \in \mathbf{E}} \|\mathbf{f}(\mathbf{e}_l) - \mathbf{f}_{\text{med}}(\mathbf{E})\|_2 \quad (4)$$
Assign the index of element $\mathbf{e}_k$ to the first matrix position, $\mathbf{M}[1,1] = k$, and remove $\mathbf{e}_k$ from $\mathbf{E}$.
2. **Transversal filling:** Fill the remaining positions of $\mathbf{M}$ following a snake scan order (left to right on one row and then right to left on the next) and according to the following assignment rule: for each new matrix position, $(i,j)$, determine its set of already assigned neighboring elements, $\boldsymbol{\mathcal{N}}_{ij}$. Neighbors include any assigned elements to the positions that are directly to the left, right, top, top-left, or top-right of $(i,j)$. The positions below (bottom-left, bottom, and bottom-right) are not considered, as they remain unfilled during the snake scanning. Compute the element-wise average of the feature vectors of these neighbors, $\mathbf{f}_{\text{avg}}(\boldsymbol{\mathcal{N}}_{ij})$. Select from set $\mathbf{E}$ the element $\mathbf{e}_k$ whose feature vector is closest to this average:
$$\mathbf{e}_k = \arg\min_{\mathbf{e}_l \in \mathbf{E}} \left\|\mathbf{f}(\mathbf{e}_l) - \mathbf{f}_{\text{avg}}(\boldsymbol{\mathcal{N}}_{ij})\right\|_2 \quad (5)$$
Assign the index of element $\mathbf{e}_k$ to $\mathbf{M}[i,j]$ and remove $\mathbf{e}_k$ from $\mathbf{E}$.
3. **Stopping criterion:** Repeat the second step until all positions of $\mathbf{M}$ are assigned and the input set is empty (i.e., $\mathbf{E} = \emptyset$).

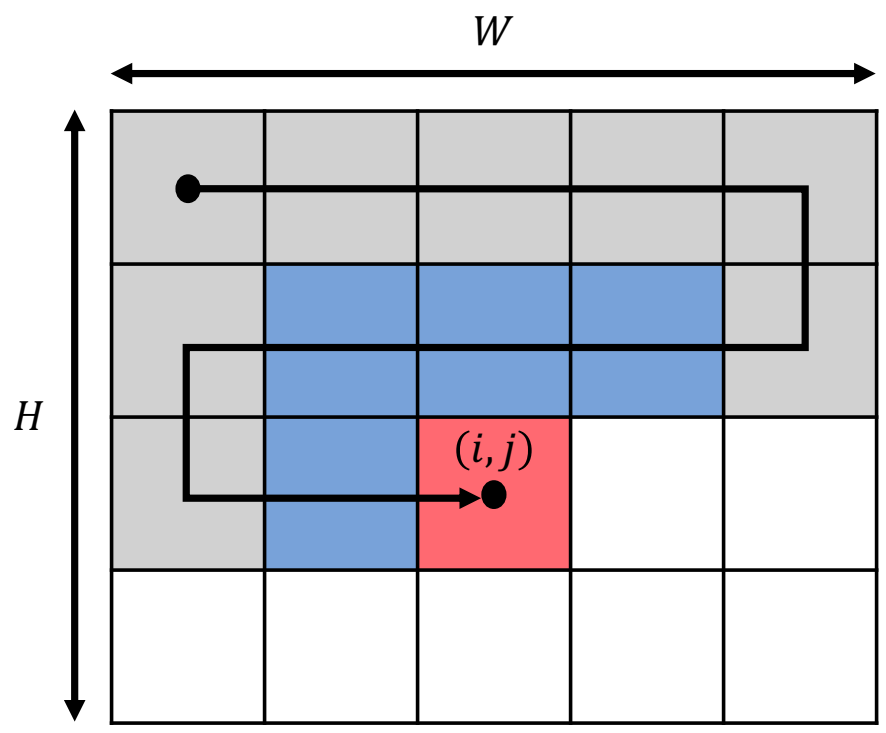

**Fig. 4:** Illustration of the NNS algorithm concept.

Each element of the NNS output matrix, $\mathbf{M}$, represents either the index of a cluster (for the cluster-to-bock assignment procedure) or the index of a Gaussian (or voxel) within a cluster (for the block filling procedure). Fig. 4 illustrates the key concept of NNS algorithm, where the red cell indicates the current matrix position, $(i,j)$, being analyzed at each iteration of the NNS algorithm, blue cells correspond to the neighboring positions considered during that iteration, $\boldsymbol{\mathcal{N}}_{ij}$, and grey cells denote positions that have already been assigned to a cluster or Gaussian (or voxel). The NNS algorithm scans the matrix in a snake order, progressively filling elements in each position considering the feature similarity with its already assigned neighbors.

The first time NNS is used on the GSICO encoder is for deciding the cluster-to-block assignment, i.e., to define the position, at the block level, of the formed fixed-size clusters. This procedure receives as input the set of $L$ fixed-size clusters, of 256 Gaussians (or voxels) each, that resulted from the clustering procedure, $\mathbf{C} = \{\mathbf{C}_1, \ldots, \mathbf{C}_L\}$, and the width, $W_{map}$, and height, $H_{map}$, of the 2D parameter maps. The output is a 2D matrix, $\mathbf{M}_{blocks}$, with dimension $W_{map}/16 \times H_{map}/16$. The cluster-to-block assignment follows two major steps:

1. **Initialization:** Compute the set of cluster centroids, $\mathbf{C_c} = \{\mathbf{c}_1, \ \ldots, \mathbf{c}_L\}$, where the centroid of each cluster is given by the element-wise average of its elements feature vectors, $\mathbf{c}_i = \mathbf{f_{avg}}(\mathbf{C}_i)$.
2. **NNS algorithm:** Run the NNS algorithm on $\mathbf{C_c}$, $\mathrm{NNS}(\mathbf{C_c},\ W_{map}/16\ , H_{map}/16)$, to obtain a 2D matrix, $\mathbf{M}_{blocks}$, where every position of $\mathbf{M}_{blocks}$ contains the index of a cluster.

After the cluster-to-block assignment, NNS is used again for block filling, i.e., to define the position, at the pixel level, of the Gaussians (or voxels) of each cluster in the respective block. This procedure receives as input the set of $L$ fixed-size clusters of 256 Gaussians (or voxels) elements, $\mathbf{C} = \{\mathbf{C}_1, \ldots, \mathbf{C}_L\}$, $\mathbf{M}_{blocks}$, and the 2D maps width, $W_{map}$, and height, $H_{map}$. The output is a 2D matrix, $\mathbf{M}_{all}$, with size $W_{map} \times H_{map}$. The block filling can be described in the steps:

1. **First block filling:** Let $\mathbf{C}_i$ denote the cluster previously assigned to the first block (specified in $\mathbf{M}_{blocks}[1,1]$). The NNS algorithm is applied to the Gaussians (or voxels) in $\mathbf{C}_i$ with a target block size of 16×16 pixels, $\mathrm{NNS}(\mathbf{C}_i,\ 16,\ 16)$. The resulting matrix is used to fill the corresponding block region in $\mathbf{M}_{all}$, the global matrix that stores the spatial assignments of all Gaussians (or voxels) across the full set of 2D parameter maps.
2. **Local block filling:** Following a block-by-block snake scan order, run $\mathrm{NNS}(\mathbf{C}_i,\ 16,\ 16)$ with the Gaussians (or voxels) corresponding to the cluster $\mathbf{C}_i$ assigned to the current block of the 2D maps according to $\mathbf{M}_{blocks}$. The resulting matrix is used to fill the corresponding block region in $\mathbf{M}_{all}$.
3. **Stopping criterion:** Repeat the second step until all positions of $\mathbf{M}_{all}$ are filled.

After cluster-to-block assignment and filling is completed, $\mathbf{M}_{all}$ is used for a final step where all Gaussian (or voxel) parameter values are mapped into the set of 2D maps according to their defined index positions. The resulting parameter maps are stored as image files, whose bit depth depends on the applied quantization (cf. Table II). All maps are stored in 8-bit PNG format, except the position maps, which require higher precision and are therefore stored as a 16-bit PNG. Fig. 5 shows a randomly organized parameter map of the *truck* scene trained (from Tanks and Temples dataset [25]) with the 3DGS model, alongside the same map produced using the proposed NNS-based GS mapping. In this example, only a subset of 2304 Gaussians was considered (for visualization purposes) and only one of the luminance SH AC parameters (one of the GS parameters used in the NNS similarity criterion for 3DGS-based inputs) is represented. Notably, the random map results in a PNG file with 1558 bytes (using only the lossless PNG compression), whereas the NNS-based map leads to 995 bytes (using the the NNS algorithm followed by lossless PNG compression). This highlights the importance of NNS in the GS parameter mapping step of the GSICO encoder.

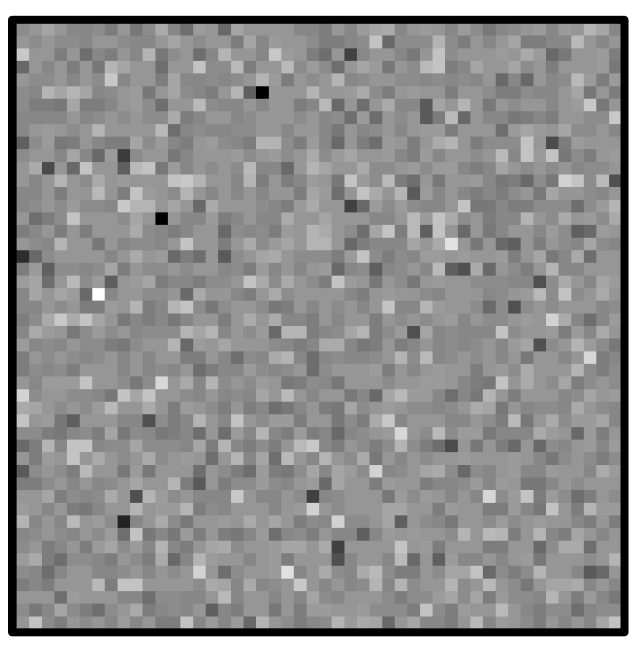

(a) Random map

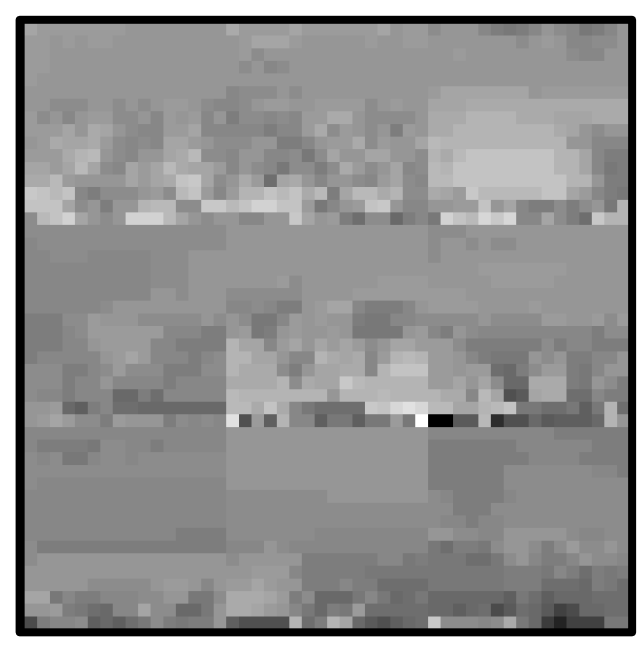

(b) NNS-based map

**Fig. 5:** Impact of the NNS-based GS mapping on the spatial organization of GS parameters.

## IV. Performance Assessment

To evaluate the GSICO performance, a comprehensive RD analysis was conducted. This section first describes the experimental setup, including the test material, rate and distortion metrics, and selected benchmarks. It then presents and discusses the experimental results, followed by an ablation study to measure the impact of each GSICO module on its overall performance.

### *A. Test conditions*

The test conditions were defined based on previous work, with the objective of ensuring reproducibility and comparability with existing solutions.

#### 1) **Test material**

The Tanks and Temples (T&T) [25], Deep Blending (DB) [26], and Mip-NeRF360 [27] were selected to ensure a comprehensive evaluation. The T&T dataset includes two complex real-world outdoor scenes, *train* and *truck*, exhibiting varied geometry and texture richness. The DB dataset contains two complex real-world indoor scenes, *drjohnson* and *playroom*, with varying lighting conditions and fine geometric details. The Mip-NeRF360 dataset comprises four real-world indoor scenes, *bonsai*, *counter*, *kitchen*, and *room*, and five real-world outdoor scenes, *bicycle*, *flowers*, *garden*, *stump*, and *treehill*, with significant variation in depth and texture. Each dataset is organized into training and test image sets. Training images are used by the GS model to reconstruct the scene (i.e., for training), while test images are reserved for rendering novel views and evaluating the performance of the model reconstruction and its compression. The two sets are disjoint, ensuring that the reported RD results are fairly obtained from viewpoints not seen during training.

TABLE III
GSICO RD RESULTS COMPARING WITH THE BASELINE MODELS

| Method | Tanks and Temples | | | | Deep Blending | | | | Mip-NeRF360 | | | |
|---|---|---|---|---|---|---|---|---|---|---|---|---|
| | PSNR↑ | SSIM↑ | LPIPS↓ | Size (MB)↓ | PSNR↑ | SSIM↑ | LPIPS↓ | Size (MB)↓ | PSNR↑ | SSIM↑ | LPIPS↓ | Size (MB)↓ |
| 3DGS | 23.88 | 0.848 | 0.176 | 306.9 | 29.27 | 0.899 | 0.260 | 223.2 | 27.52 | 0.815 | 0.214 | 556.2 |
| Scaffold-GS | 24.13 | 0.854 | 0.176 | 80.0 | 30.33 | 0.909 | 0.253 | 57.4 | 27.73 | 0.815 | 0.221 | 187.6 |
| GSICO (w/ 3DGS) | 23.50 | 0.837 | 0.189 | 15.0 | 29.15 | 0.890 | 0.265 | 11.3 | 26.70 | 0.798 | 0.232 | 27.6 |
| GSICO (w/ Scaffold-GS) | 24.03 | 0.849 | 0.179 | 11.2 | 30.18 | 0.907 | 0.256 | 6.3 | 27.27 | 0.800 | 0.232 | 20.8 |

2) **Rate and quality metrics**

The full bitstream size, expressed in megabytes (MB), was adopted as the rate measure. For quality evaluation, five widely recognized full-reference image quality metrics were considered. DISTS [28] and FSIM [29] metrics, both shown to exhibit high correlation with human perception [30], were selected for the RD characterization of GSICO and the ablation study. DISTS combines deep feature representations with texture and structure similarity, while FSIM computes a normalized similarity measure based on salient image features. To facilitate RD curve visualization, both DISTS and FSIM results were transformed into decibel (dB) scales using $-20\log_{10}(m)$ (for DISTS) and $-20\log_{10}(1-m)$ (for FSIM), where $m$ denotes the original metric value. Additionally, PSNR, SSIM [31], and LPIPS (VGG) [32] were included to enable direct comparison with existing GS compression methods, as these three metrics are consistently reported in prior works. PSNR evaluates per-pixel fidelity, SSIM measures structural consistency and luminance-contrast similarity, and LPIPS assesses perceptual similarity based on deep NN features. All metrics were computed in the RGB color space, except FSIM which operates in the LAB color space. SSIM and FSIM range from 0 (lowest quality) to 1 (highest quality) while LPIPS and DISTS ranges from 0 (highest quality) to 1 (lowest quality). For all metrics, the reference corresponds to the test images provided in the datasets, in line with common practice in radiance field evaluation [33]. Metric values were first computed per test image (of a scene) and then averaged to obtain a single score per scene. The score of a dataset corresponds to the average of the scores of its scenes.

3) **Selected benchmarks**

GSICO performance was compared against the baseline models (3DGS [2] and Scaffold-GS [4]) and several state-of-the-art GS compression methods, selected for their diversity in terms of technical approach and competitive performance. These include: Compact3D [19], employing vector quantization, compressing the codebooks via index sorting and run-length encoding; Reduced3DGS [12], implementing resolution-aware Gaussian pruning, adapting SH coefficient maximum degree per Gaussian, and applying adaptive visual importance-aware quantization; LightGaussian [13], pruning Gaussians based on scene spatial coverage, distilling SH coefficients to lower degrees, and applying vector quantization; CompGS [14], using a learnable mask to remove Gaussian with low visual impact, employing vector quantization, and building an entropy model based on GS parameter hyperpriors and inter-parameter priors; EAGLES [34], integrating entropy-aware coding with spatial Gaussian attributes prediction; SOG [15], pruning low-opacity Gaussians and implementing an image-based Gaussian parameters compression; CodecGS [16], using vector quantization built upon tri-plane image representation; RDO Gaussian [35], applying RD optimization to allocate bit depths dynamically across parameters; and FCGS [36], focusing on streamlined data structures and transforms optimized for fast coding. All benchmarks were evaluated using the parameter settings recommended by the authors in their works to ensure a fair comparison. Although recent GS methods such as HAC [17], HAC++ [18], HEMGS [37], and ContextGS [38] have reported competitive GS compression performances, they were excluded from the comparison: HEMGS due to the lack of a publicly available implementation, while the remaining methods did not provide a functional decoder at the time of this work.

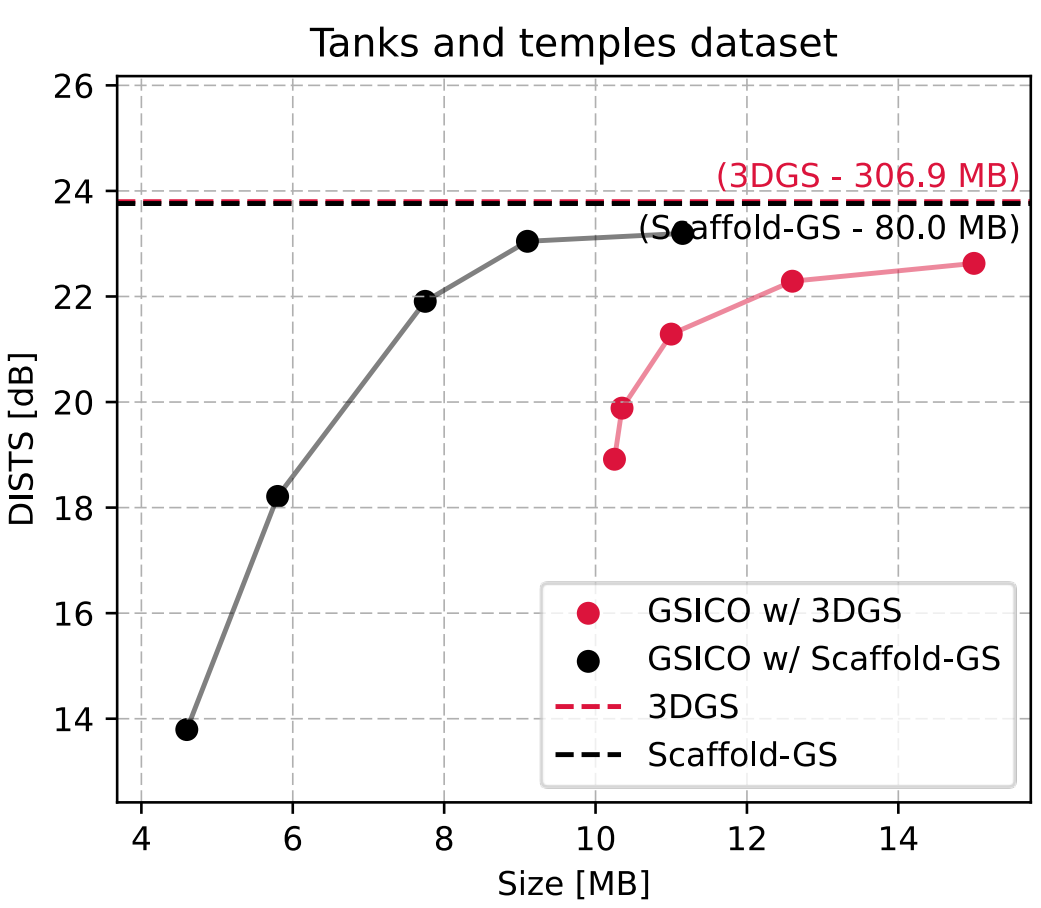

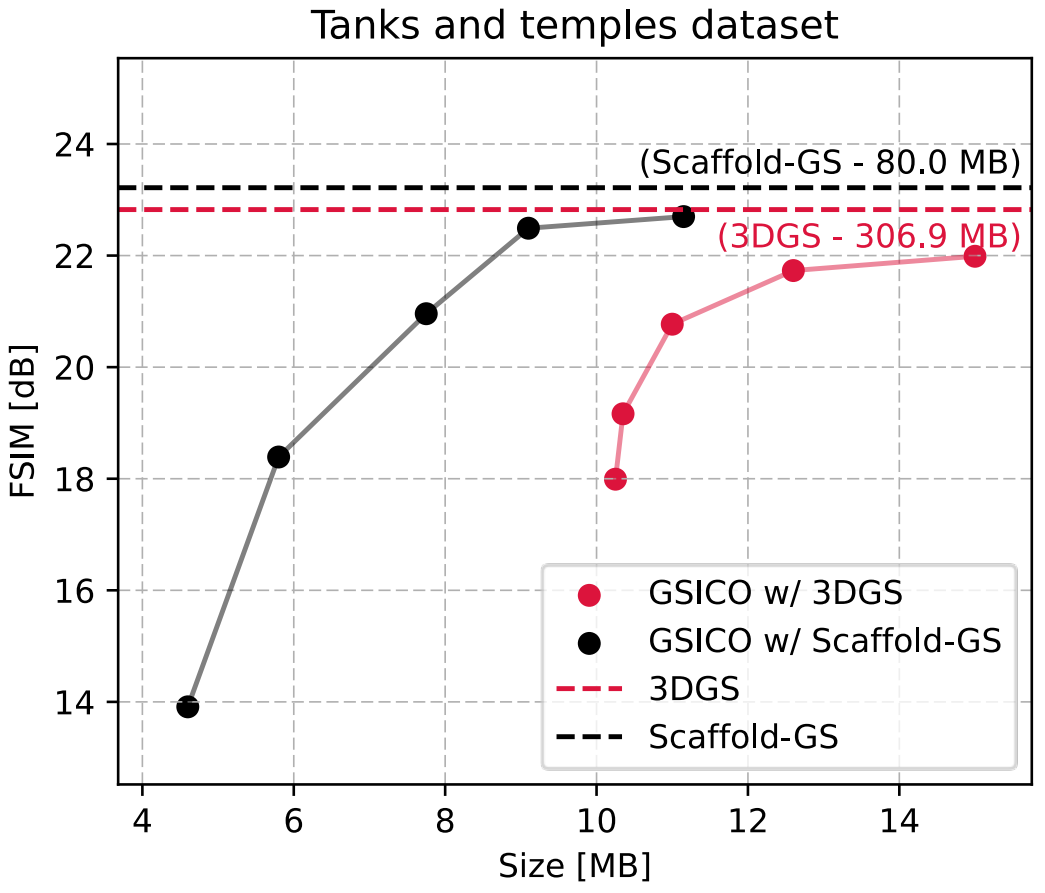

**Fig. 6:** RD performance of GSICO codec and the 3DGS and Scaffold-GS baseline models.

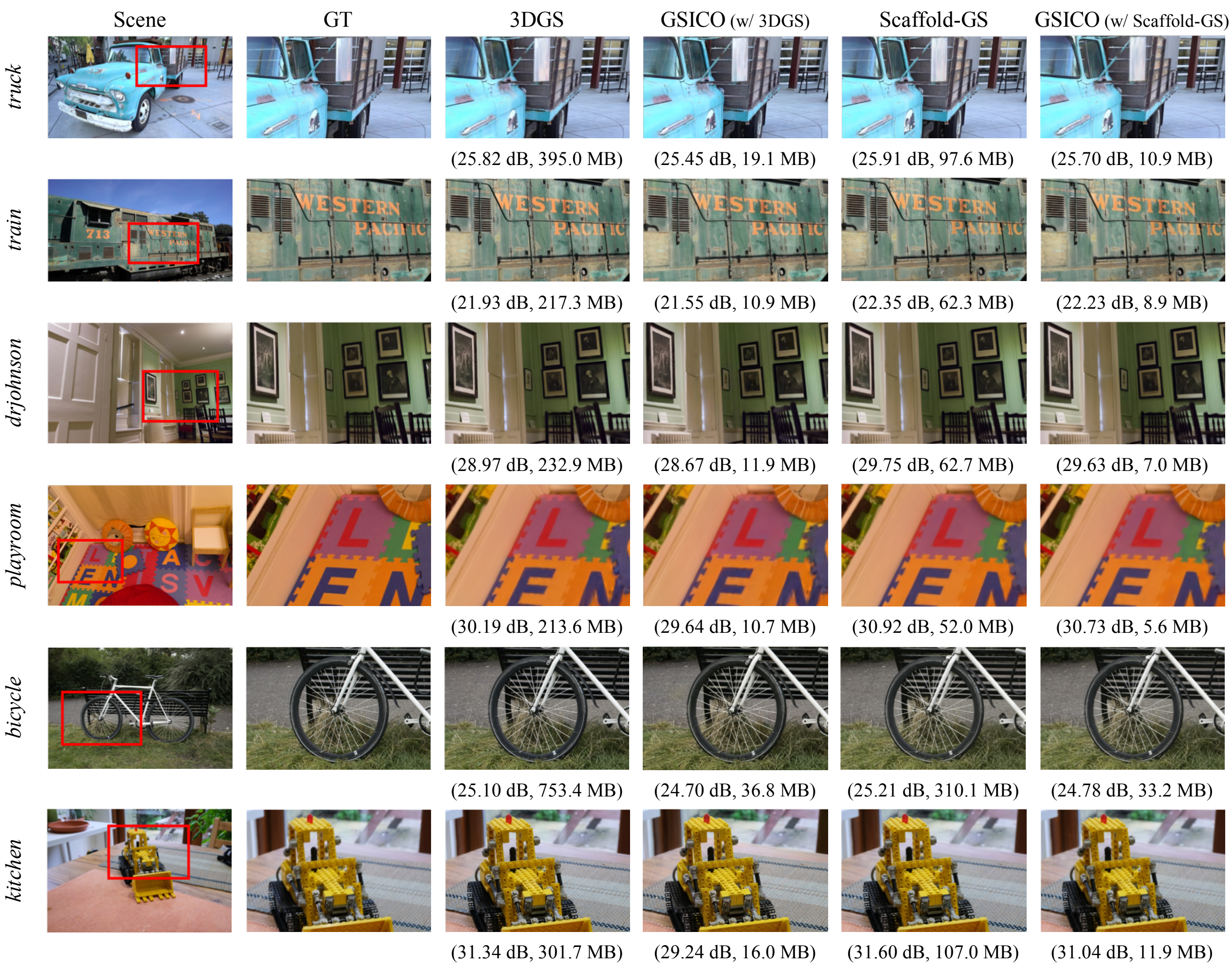

**Fig. 7:** Qualitative comparison between ground-truth (GT) images, baseline model renderings, and GSICO-compressed renderings for selected scenes across all datasets, shown with the corresponding average PSNR values and bitstream sizes.

## *B. Experimental Results*

This section presents the RD performance of GSICO, compares its compression gains with baseline models and GS compression state-of-the-art benchmarks, and reports an ablation study on the GSICO components.

### 1) **RD characterization of the proposed GSICO codec**

To fully characterize the RD performance of the proposed GSICO codec, five operating points were defined for each input representation model (cf. Section III.A). Fig. 6 shows the GSICO performance results obtained for the T&T dataset. The RD curves exhibit excellent rate scalability and stability across operating points. For both 3DGS-based and Scaffold-GS-based configurations, GISCO covers a wide operating range, from highly compressed to near-lossless regimes (i.e., same quality as the baseline without any compression), while maintaining smooth and monotonic quality progression. Notably, no abrupt quality drops are observed between adjacent operating points, demonstrating the robustness and consistency of the quantization strategy across bitrate levels.

### 2) **Improvements over the GS baseline models**

A comparison between GSICO and the corresponding uncompressed baseline models is shown in Table III, considering only the highest-quality GSICO operating point, which closely matches the quality of baseline models. These results demonstrate that GSICO achieves substantial reductions in model size while preserving high rendering quality.

For 3DGS-based inputs, GSICO reduces the model size by an average factor of 20.2× (from 306.6 MB to 15.0 MB on T&T, from 223.2 MB to 11.3 MB on DB, and from 556.2 MB to 27.6 MB on Mip-NeRF360). The quality degradation is minimal: average PSNR drops are limited to 0.45 dB (from 23.88 dB to 23.50 dB on T&T, from 29.27 dB to 29.15 dB on DB, and from 27.52 dB to 26.70 dB on Mip-NeRF360), SSIM variations remain below 0.01, and LPIPS degradation does not exceed 2%. These results indicate that GSICO efficiently compresses GS parameters without introducing significant perceptual degradations.

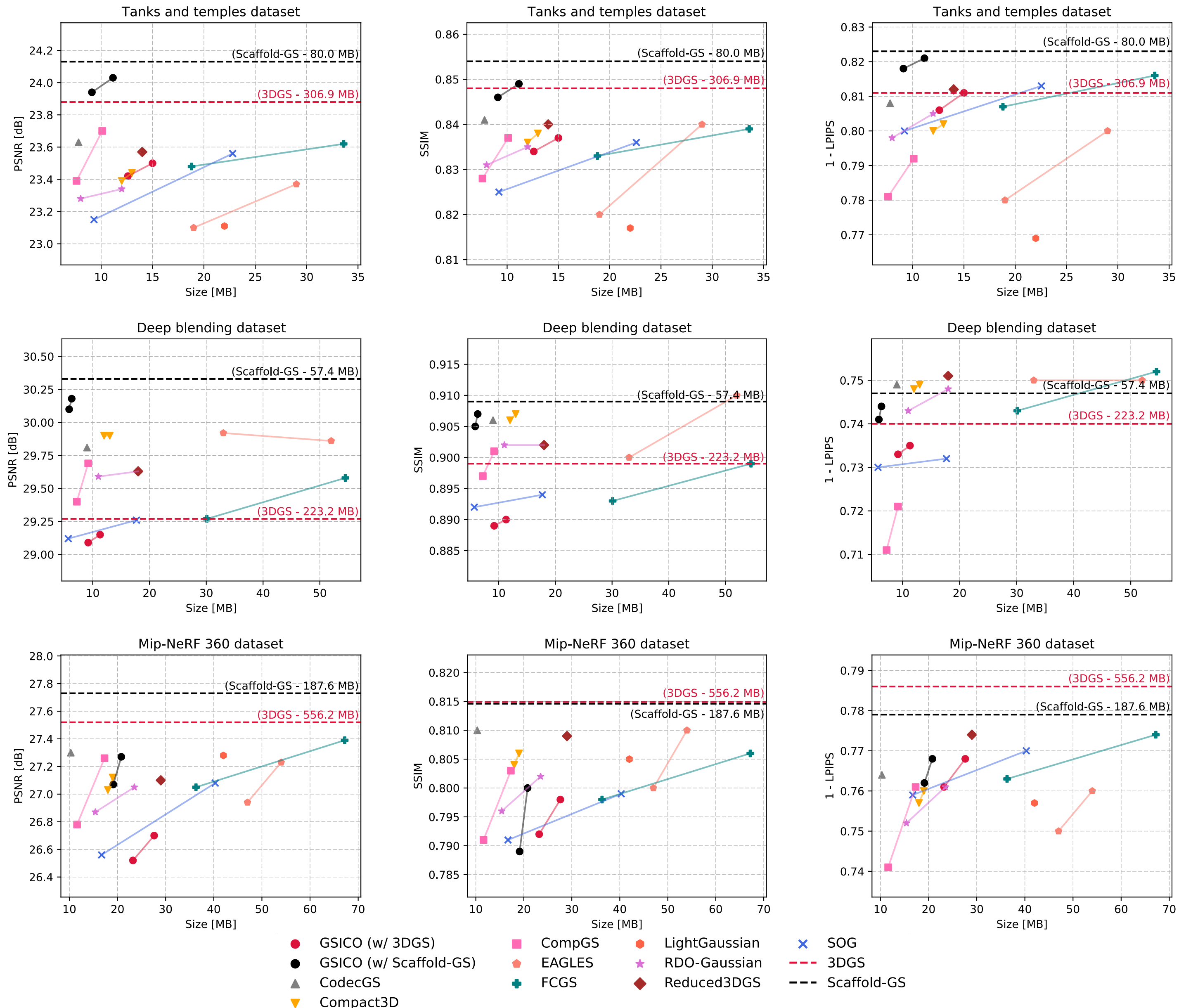

**Fig. 8:** GSICO RD curves compared with selected benchmarks.

For Scaffold-GS inputs, GSICO achieves an average compression factor of 8.5× (from 80.0 MB to 11.2 MB on T&T, from 57.4 MB to 6.3 MB on DB, and from 187.6 MB to 20.8 MB on Mip-NeRF360). Average PSNR reduction is 0.38 dB (from 24.13 dB to 24.03 dB on T&T, from 30.33 dB to 30.18 dB on DB, and from 27.73 dB to 27.27 dB on Mip-NeRF360), SSIM variations remain below 0.01, and LPIPS values do not degrade more than 1.5%, further supporting the efficiency of the codec in preserving rendering quality. Although lower than the compression gains observed for 3DGS, these reductions are particularly significant given the already more compact nature of Scaffold-GS representation.

A qualitative evaluation of the proposed codec was also conducted on representative scenes from all the selected datasets. Fig. 7 presents visual comparisons (after synthesis) between ground-truth (GT) images, uncompressed GS baseline models, and GSICO. Even under substantial compression, GSICO preserves the quality of the rendered views, with no noticeable degradations such as blurring, color shifts or structural distortions. Both 3DGS-based and Scaffold-GS-based configurations maintain visual consistency across complex indoor and outdoor scenes, reinforcing GSICO's ability to significantly reduce model size while preserving perceptual quality.

3) **Benchmark evaluation**

For comparison with state-of-the-art GS compression methods, only the two highest-quality GSICO operating points were considered, ensuring a fair and balanced comparison with benchmarks that typically operate over a narrower RD range. Fig. 8 illustrates the RD performance across all datasets.

On the T&T dataset, the Scaffold-GS-based GSICO configuration emerged as the top-performing solution, with the best trade-off between quality and compression efficiency. At a rate of 10 MB, GSICO outperforms CompGS by approximately 0.3 dB in PSNR. The GSICO 3DGS-based configuration, although less competitive than its Scaffold-GS counterpart, still outperforms the image-based compression method SOG, requiring approximately 5.9 MB less bitrate at a PSNR of 23.5 dB.

On the DB dataset, the Scaffold-GS-based GSICO configuration again delivers the best performance, surpassing all benchmarks. Relative to the strongest competing solutions, Compact3D and CodecGS, at a SSIM of 0.906, GSICO requires approximately 5.85 MB and 2.85 MB less rate, respectively. Although the GSICO 3DGS-based configuration is less competitive in this case, it still surpasses SOG in terms of LPIPS, the quality metric with highest correlation with human perception from the three selected quality metrics [30].

On the Mip-NeRF360 dataset, which contains large-scale and geometrically complex indoor and outdoor scenes, GSICO maintains strong performance. The Scaffold-GS-based configuration achieves competitive quality, particularly in terms of LPIPS: at 19.1 MB (corresponding to its low-rate configuration), it achieves a quality level similar to the high-rate configurations of both Compact3D and RDO-Gaussian. GSICO 3DGS-based configuration, while achieving more modest compression gains, it still provides a superior RD performance compared to SOG for both SSIM and LPIPS curves. The challenging nature of the Mip-NeRF360 dataset, where GS methods tend to produce a higher number of Gaussians than in other datasets, makes compression particularly difficult. In this case, joint GS training and compression methods tend to perform better.

Across all datasets, the RD performances depicted in Fig. 8 confirm that GSICO consistently has the best RD tradeoff, with its Scaffold-GS-based configuration emerging as the best solution. Overall, GSICO distinguishes itself from prior work through its training-free and model-agnostic design, ensuring immediate applicability across diverse GS models and thus supporting practical deployment in real-world multimedia systems.

4) **Ablation study**

The ablation study selectively disabled individual GSICO components to provide insight into their respective contribution to the overall system behavior. RD performance is evaluated for each supported baseline model (3DGS and Scaffold-GS) using the T&T scenes. The study begins to assess the impact of the first GSICO component, namely pruning. Since pruning is required in GSICO to enable a rectangular 2D map arrangement of GS parameters, its impact is evaluated relative to the uncompressed model. Subsequently, the full GSICO configuration is used as the reference to analyze the contributions of the NNS-based mapping and the JPEG XL coding components. Specifically, GSICO is compared against: *i*) a variant employing random GS mapping, which emulates the absence of the proposed NNS-based mapping, and *ii*) a variant in which JPEG XL coding is disabled and the 2D maps are instead encoded using a default PNG format. Finally, to gain insight into the impact of the quantization component (whose behavior depends on the 2D map creation) a third variant is introduced in which both NNS-based mapping and JPEG XL coding are disabled. This variant employs random mapping for 2D map creation and relies primarily on the GS parameter quantization.

TABLE IV
ABLATION STUDY RESULTS

(a) 3DGS-based input files

| Technique | FSIM↑ | DISTS↓ | Size [MB]↓ |
|---|---|---|---|
| **3DGS (no compression)** | **0.925** | **0.071** | **306.2** |
| 3DGS (w/ pruning) | 0.925 | 0.071 | 306.1 |
| **GSICO** | **0.921** | **0.074** | **15.0** |
| GSICO (w/o NNS-based mapping) | 0.920 | 0.075 | 21.6 |
| GSICO (w/o JPEG XL coding) | 0.922 | 0.072 | 21.3 |
| GSICO (w/o NNS and JPEG XL) | 0.922 | 0.072 | 29.9 |

(b) Scaffold-GS-based input files

| Technique | FSIM↑ | DISTS↓ | Size [MB]↓ |
|---|---|---|---|
| **Scaffold-GS (no compression)** | **0.929** | **0.068** | **80.0** |
| Scaffold-GS (w/ pruning) | 0.929 | 0.068 | 79.9 |
| **GSICO** | **0.926** | **0.070** | **9.9** |
| GSICO (w/o NNS-based mapping) | 0.926 | 0.070 | 10.7 |
| GSICO (w/o JPEG XL coding) | 0.926 | 0.070 | 11.3 |
| GSICO (w/o NNS and JPEG XL) | 0.926 | 0.070 | 12.5 |

For both baseline model configurations, the pruning component has a negligible impact on RD performance, as expected, since its role is solely to impose the structural constraints required to arrange GS parameters into 2D maps. Regarding the GS mapping, the proposed NNS-based strategy consistently improves compression efficiency compared to random mapping, especially for the 3DGS-based configuration. This behavior can be attributed to the higher parameter variability exhibited by Scaffold-GS, which limits the exploitation of spatial coherence, leading to less smooth NNS-based maps. Regarding the image coding, JPEG XL outperforms PNG in terms of compression efficiency for the evaluated variant. In the 3DGS-based configuration, the lossy JPEG XL coding applied to some maps introduces a residual quality degradation, whereas the Scaffold-GS-based configuration employs fully lossless JPEG XL coding and therefore does not exhibit any quality degradation. Finally, quantization yields the most significant RD gains, particularly in the 3DGS-based configuration, where the combination of SH color space conversion and the removal of the chrominance AC coefficients lead to a significant reduction in model size.

## V. FINAL REMARKS AND FUTURE WORK

The main objective of this work was to address the challenge of efficiently compressing GS models for practical deployment in real-world multimedia scenarios. GS methods, notably 3DGS and Scaffold-GS, have emerged as leading representations for radiance fields due to their favorable trade-off between rendering quality and computational efficiency. However, their high storage requirements limit their applicability in bandwidth-constrained or storage-limited environments. This paper addresses this limitation by introducing GSICO, an efficient, versatile, and post-training GS codec. GSICO maps GS parameters into structured 2D images, which are subsequently compressed using a standard image codec. A novel 2D GS parameter mapping strategy is proposed to enhance spatial coherence within these images, thereby significantly improving their compressibility.

Comprehensive experiments across three widely used radiance field datasets demonstrates the effectiveness of the proposed approach. When applied to 3DGS-based models, GSICO achieves an average compression factor of 20.2× with negligible perceptual degradation, while for Scaffold-GS-based models it yields an average reduction of 8.5×. RD performance evaluations further show that the GSICO Scaffold-GS-based configuration consistently outperforms existing benchmarks. Unlike most existing GS compression approaches, which tightly integrate compression mechanisms into the training process and therefore require access to the original training data, GSICO operates entirely as a post-training codec. This design choice enables immediate adoption in practical multimedia scenarios where training data are unavailable. Moreover, GSICO's compatibility with both 3DGS-based and Scaffold-GS-based representations enhances its versatility, ensuring applicability across the two most representative GS model baselines.

Future work will focus on three main directions: *i)* implementing a learning-based quantization strategy, where quantization step sizes are adaptively determined based on both the statistical properties and the perceptual relevance of each GS parameter; *ii)* while JPEG XL was selected for its outstanding performance in coding structured parameter images, emerging learning-based image codecs, such as JPEG AI [39], offer promising opportunities for even greater efficiency, provided they are properly fine-tuned to the statistical characteristics of GS parameters; and *iii)* extending the proposed codec framework to support dynamic GS models represents another compelling research direction.

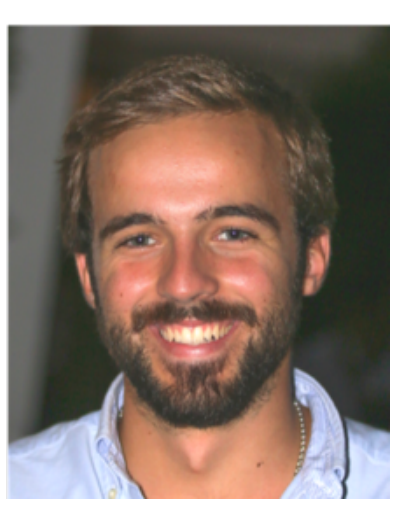

**Pedro Martin** (Member, IEEE) received the B.S. and M.S. degrees in electrical and computer engineering from Instituto Superior Técnico (IST), University of Lisbon (UL), Portugal, in 2019 and 2021, respectively. He is currently pursuing the Ph.D. degree with the Department of Electrical and Computer Engineering, IST/UL. He has been a Researcher with Instituto de Telecomunicações, since 2021, and a Teaching Assistant with the Department of Electrical and Computer Engineering, IST/UL, since 2020. His main research interests include visual quality assessment and coding, with a particular focus on neural radiance fields.

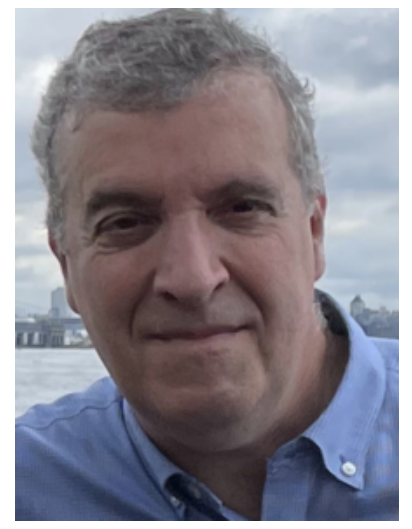

**António Rodrigues** (Member, IEEE) received the B.S. and M.S. degrees in electrical and computer engineering from Instituto Superior Técnico (IST), Technical University of Lisbon, Lisbon, Portugal, in 1985 and 1989, respectively, and the Ph.D. degree from the Catholic University of Louvain, Louvain-la-Neuve, Belgium, in 1997. Since 1985, he has been with the Department of Electrical and Computer Engineering, IST, where he is currently an Associate Professor. He is also a Senior Research Member of Instituto de Telecomunicações, Lisbon. His current research interests include mobile and satellite communications, wireless networks, modulation, coding, and multiple access techniques.

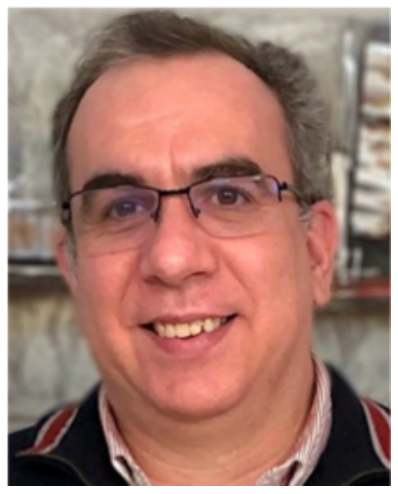

**João Ascenso** (Senior Member, IEEE) received the E.E., M.Sc., and Ph.D. degrees in electrical and computer engineering from Instituto Superior Técnico (IST), Universidade Técnica de Lisboa, Lisbon, Portugal, in 1999, 2003, and 2010, respectively. He is currently an Associate Professor with the Department of Electrical and Computer Engineering, IST, and a member of Instituto de Telecomunicações.

He has authored more than 100 papers in international conferences. His current research interests include visual coding, quality assessment, light field and point cloud processing, and indexing and searching of multimedia content. He was an Associate Editor of IEEE TRANSACTIONS ON IMAGE PROCESSING and IEEE TRANSACTIONS ON MULTIMEDIA.

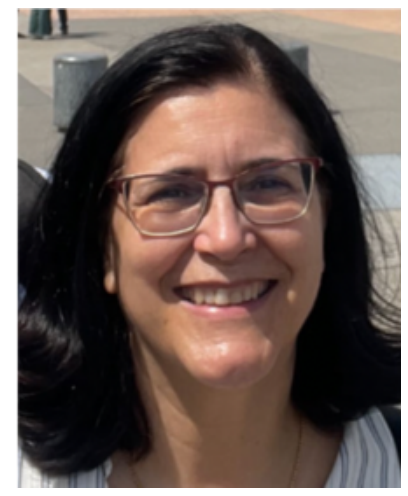

**Maria Paula Queluz** received the B.S. and M.S. degrees in electrical and computer engineering from Instituto Superior Técnico (IST), University of Lisbon, Portugal, and the Ph.D. degree from the Catholic University of Louvain, Louvain-la-Neuve, Belgium. She is currently an Associate Professor with the Department of Electrical and Computer Engineering, IST, and a Senior Research Member of Instituto de Telecomunicações, Lisbon, Portugal. Her main scientific and research interests include image/video quality assessment, image/video processing, and wireless communications.